\documentclass{article}
\usepackage{amsmath,amsfonts}
\def\imath{i}
\def\bra{\langle}
\def\ket{\rangle}
\newcommand{\R}{\mathbb{R}}

\def\cN{{\mathcal N}}
\def\cH{{\mathcal H}}             
\def\cD{{\mathcal D}}             

\def\eg{e.g.\,}
\def\d{\partial}
\def\bra{\langle}
\def\ket{\rangle}

\def\bvec#1{\mathbf{#1}}
\def\dk#1#2{\frac{ d^{#2}{#1} }{ (2\pi)^{#2} }} 
\def\da#1#2{\frac{ d{#1}}{{#1}^{{#2}+1}}}
\begin{document}
\title{$\phi^4$-Field theory on a Lie group}
\author{M.V.Altaisky \\
$^1${\em Joint Institute for Nuclear Research, Dubna, 141980, Russia} \\
$^2${\em Space  Research Institute, Profsoyuznaya 84/32, Moscow, 117810, 
Russia} \\ e-mail: altaisky@mx.iki.rssi.ru}
\date{Revised: Sep 13, 2000}
\maketitle
\begin{abstract}
The $\phi^4$ field model is generalized to the case when 
the field $\phi(x)$ is defined on a Lie group: 
$S[\phi]=\int_{x\in G} L[\phi(x)] d\mu(x)$,  
$d\mu(x)$ is the left-invariant measure on a locally compact group $G$. 
For the particular case of the affine group $G:x'=ax+b,a\in\R_+, x,b \in \R^n$ the Feynman 
perturbation expansion for the Green functions is shown to have no 
ultra-violet divergences for certain 
choice of $\lambda(a) \sim a^\nu$. 
\end{abstract}
\section{Introduction}
The ultra-violet (UV) divergences appearing in quantum field theory at small distances 
(high momentum $\Lambda\to\infty$) are well known to be 
intimately related to the properties of the theory with respect to 
the group of scale transformations. For a wide class of theories, 
known as {\em multiplicatively renormalizable} theories, the problem 
can be essentially simplified by the scale transformation of fields
($\phi$) and coupling constants ($g$)  
$$\phi_R = Z^{-1}_\phi \phi,\quad g = g_0 Z^{-1}_g.$$
The renormalized Green functions 
$$
G_n^R(x_1,\ldots,x_n;g,\Lambda) = Z_\phi^{-n}G_n(x_1,\ldots,x_n;g_0,\Lambda)
$$
become finite in the limit $\Lambda\to\infty$, with all divergences 
hidden in infinite renormalization constants 
$Z_\phi(g,\Lambda),Z_g(g,\Lambda)$. 

The independence of physical results on scale transformations 
\begin{equation} 
\Lambda'=e^l \Lambda, \quad x' = e^{-l}x
\label{sct}
\end{equation}
is known as renormalization group (RNG) equation. 

The modern quantum field theory has become inconceivable without 
RNG methods. Most of the results obtained phase transitions, 
quantum electrodynamics, quantum chromodynamics etc. are direct 
consequences of RNG methods. Therefore, we may have a temptation 
to base the theory on some kind of covariance with respect to 
scale transformations \eqref{sct} from very beginning, not after 
facing the UV divergences problem. 

The best way to study any physical system is to choose a functional 
basis with the symmetry properties as close to the symmetry of the 
system as possible. For this reason we choose the spherical functions 
to study the hydrogen atom, and for the same reason we use plane waves 
to describe a particle moving in homogeneous space. Of course it is 
also possible to apply plane waves to $SO_3$ symmetrical problem, but 
one can hardly expect any use of it. 

It is important what is implied by ``the symmetry of the problem''. 
We assume that the system is described by a set of complex-valued 
functions $\phi^\alpha$ defined on a manifold $\mathcal{M}$,  
 $\phi^\alpha := \phi^\alpha(x), x \in \mathcal{M}$.
A system is said to have a symmetry group $G$ if the action of 
the group $G$ on the independent variables ({\em coordinates}) 
and dependent variables ({\em fields})
$$
x \to x' = \hat T x, \quad 
\phi^\alpha(x) \to {\phi'}^\alpha(x') = {\hat M}^\alpha_\beta \phi^\beta(x), 
$$
where $\hat T$ and $\hat M$ are operators, does not change the action 
functional
(or any other functional which is believed to determine the dynamics of 
the system). 

If the transformation group does not affect the fields themselves 
(${\hat M}^\alpha_\beta\equiv\hat 1$), but only coordinates
$$
\phi^\alpha(x) \to {\phi'}^\alpha(x') = \phi^\alpha({\hat T}^{-1}x'),
$$
the field $\phi^\alpha$ is called a scalar with respect to the 
transformation group $G$. 

The most important group of transformations used in physics is the 
Poincare group ${x_\mu}' = \Lambda_\mu^\nu x_\nu + b_\mu$. The wave 
functions of elementary particles - electrons, photons, quarks etc. - 
are not Poincare scalars. They have nontrivial transformation properties 
under Lorenz rotations $\Lambda_\mu^\nu$ and are classified according 
to their spin. However, it is possible to consider certain simplistic 
models with scalar fields, which do have, or may have physical implications 
for real systems. One of the most known models is the scalar theory of 
critical behavior, where magnetization $\phi(x)$ is considered as a 
function of the coordinate in Euclidean space. The scalar field theory 
was application point of the Wilson renormalization group awarded by a 
Nobel prize in 1982. The scalar field theory in Euclidean space 
is an analytical continuation ($\tau=it$) of a field theory in 
Minkovski space, and is receiving a lot of attention.

In this paper we restrict ourselves to the theory of complex-valued 
scalar field. Usually, the scalar field theory is defined on the 
$n$-dimensional Euclidean space $\R^n$, which is isomorphic to the group 
of translations 
\begin{equation}
x' = x+b,\quad x,b \in \R^n
\label{tg}
\end{equation}
The representation of the translation group \eqref{tg} on the space 
of square-integrable functions is given by 
$
U(b)\phi(x) = \phi(x-b)
$.
The unitary representation of the translation group is defined 
on the space of periodic functions 
$$U(b) e^{-imx} = e^{imb} e^{-imx},\quad  U(-b) = U^*(b).$$
Thus it is possible to decompose a function $\phi(x)$ with 
respect to the representations of translation group $G$
\begin{equation}
\phi(x) = \int_G e^{ixb} \hat \phi(b) db.
\label{ft} 
\end{equation}
This is Fourier decomposition.
Similarly, a function may be decomposed with respect to $SO_3$ rotations, 
Poincare group \cite{Klauder} and other groups. 

Since the concept of the group is just more general than the concept 
of the Euclidean or Minkovski space, a question naturally arises: {\em For 
what groups it is physically meaningful to construct a decomposition 
like \eqref{ft} and use may we have of it in field theoretic calculations}?

From physical point of view, the coordinates ($x$) can not be measured with 
arbitrary high precession, and it seems more reasonable to speak about 
the values of the fields  $\phi^\alpha$ measured at a position $x$ with 
the finite resolution $\Delta x$. {\em The claim of the present paper is 
that an adequate description of this situation, which inherits the 
RNG ideas, can be achieved if 
we use an analog of \eqref{ft} decomposition on the base of the affine 
group}
\begin{equation}
x' = a x + b
\label{ag}, 
\end{equation} 
where, as it will be seen later, $a$ can be understood as resolution and 
$b$ as a coordinate. 

The goal of the present paper is to construct a $\phi^4$ model where the 
scalar field $\phi(a,b)$ is defined on the affine group (in the sense 
that $a,b$ a coordinates on the affine group \eqref{ag}) and to link 
the new model with renormalization properties of the known $\phi^4$ model 
in $\R^n$. 

The paper is organized as follows. In {\em section 2} we review the basic 
formalism of $\phi^4$ theory in $\R^n$. In {\em section 3} we remind the 
technique of wavelet transform with respect to a locally compact Lie groups. 
In {\em section 4} the $\phi^4$ theory on the affine group is presented.

\section{$\phi^4$ Field theory} 
The scalar field theory with the forth power interaction 
$\frac{\lambda}{4!}\phi^4(x)$ defined on Euclidean space $x\in\R^n$ is 
one of the most instructive models any textbook in field theory starts with, 
see \eg \cite{Ramond1981}. Often called a Ginsburg-Landau model for its 
ferromagnetic counterpart, the model describes a quantum field 
with the (Euclidean) action 
\begin{equation}
S[\phi] =\int d^n x 
\frac{1}{2}(\d_\mu\phi)^2+ \frac{m^2}{2}\phi^2 + \frac{\lambda}{4!}\phi^4(x).
\label{f4l}
\end{equation}
in $n$-dimensional Euclidean space. Alternatively, the theory of quantum 
field in Euclidean space is equivalent to the theory of classical fluctuating 
field with the probability measure $\cD P = e^{-S[\phi]}\cD\phi$. In this case 
$m^2$ is the deviation from critical temperature $m^2 = |T-T_c|$, and 
$\lambda$ is the fluctuation interaction strength. To some extent, the 
$\phi^4$ model considered in this way describes a second type phase 
transition at zero external field in any system with one-component order 
parameter $\phi=\phi(x)$ and symmetry $\phi \to -\phi$.   

 The Green functions (correlation functions) 
\begin{equation}
G_m(x_1,\ldots x_m) \equiv \bra \phi(x_1)\ldots\phi(x_m)\ket = \frac{1}{W_E[J]} 
\left. \frac{\delta^n}{\delta J(x_1)\ldots\delta J(x_m)}\right|_{J=0} W_E[J]
\label{gfm}
\end{equation}
are evaluated as functional derivatives 
of the generating functional
\begin{equation}
W_E[J] = \cN \int \cD\phi \exp\left[-S[\phi(x)] + \int  d^nx J(x)\phi(x) \right] 
\label{gf}.
\end{equation}
(The formal constant $\cN$ associated with functional measure 
$\cD\phi$ dropped hereafter.)
 
 The straightforward way to calculate $G_m$ is to factorize the 
interaction part $V(\phi)$ of 
generation functional \eqref{gf} in the form
\begin{equation}
W_E[J] = \exp\left[-V\left(\frac{\delta}{\delta J}\right) \right] W_0[J],
\end{equation} 
where 
\begin{equation}
W_0[J] = \int \cD\phi \exp\left(J\phi - \frac{1}{2}\phi D\phi  \right) = 
\exp\left(-\frac{1}{2}JD^{-1}J \right), \quad 
D = -\d^2 +m^2\label{w0}
\end{equation}
is the free part of the generating functional.  
The perturbation expansion is then evaluated in $k$-space, where 
$$
{\widehat D}^{-1}(k) = \frac{1}{k^2+m^2}.
$$
The perturbative calculation of the correlation functions in $n>2$ dimensions 
suffers ultra-violet divergence starting from one-loop approximation
\begin{equation}
I_1(n) = -\frac{\lambda}{2} \int \dk{k}{n}\frac{1}{k^2+m^2}. 
\label{l1i}
\end{equation}
The $\phi^4$ is renormalizable, i.e. the divergences can be eliminated by 
renormalization of the fields and parameters 
\begin{equation}
\phi = Z_\phi \phi_R, \quad \lambda_0 = \lambda m^{2\epsilon}Z_\lambda,\quad 
m_0^2 = m^2 Z_m,
\end{equation} 
where all divergences are hidden in infinite renormalization 
constants $Z_\phi, Z_\lambda, Z_m$. 

Technically, the elimination of divergences is related to the evaluation 
of the loop integral \eqref{l1i} in the spherical domain in $k$-space 
limited from 
above $|k|<\Lambda$, with substitution of fixed coupling 
constant $\lambda_0$ to running coupling constant $\lambda=\lambda(\Lambda)$.
In the case of $\phi^4$ theory in the dimension 
$n=4-\epsilon$, the renormalization, as it was 
shown by K.Wilson \cite{Wilson1973}, leads to the exact scaling of the coupling constant 
\begin{equation}
\lambda(\Lambda) = \lambda_0 \Lambda^\epsilon
\label{ws}
\end{equation}
at the limit of infinite cutoff momentum $\Lambda\to\infty$. 
Similar type scaling takes place for other types of 4-th power 
interactions, say for Fermi interaction 
$G_0 (\bar\psi \psi)^2$. 

If we believe, that the power-law dependence of coupling constant 
on the cutoff 
momentum, really means the dependence of interaction strength on the 
scale $a=\Lambda^{-1}$, rather than a pure mathematical trick, we should 
find a way to incorporate this dependence at the level of the basic model, 
rather than at technical level. Doing so, after reviewing some necessary 
facts from group representation theory in next section, we will use 
the decomposition (often referred to as {\em wavelet transform}) with 
respect to the affine group  for this purpose.  

\section{Partition of the unity} 
From the group theory 
point of view, 
the reformulation of the theory from the coordinate representation 
$\phi(x)$ to the momentum representation $\hat\phi(k)$ by means  of 
Fourier transform \eqref{ft}, is only a particular case of decomposition 
of a function with respect to representation of a Lie group $G$. 
$G: x'=x+b$ for the case of Fourier transform, but other groups 
may be used as well, depending on the physics of a particular problem.

Let us remind briefly how the decomposition with respect to the given  
representation of a Lie group is performed \cite{Carey76,DM1976}. 
Let $\cH$ be a Hilbert space. Let $U(g)$ be a square-integrable 
representation of a locally-compact Lie group $G$ acting transitively 
on $\cH$, $\forall \phi\in\cH,g\in G : U(g)\phi \in \cH$. Let there 
exist such a vector  $|\psi\ket \in \cH$, that satisfies  
{\em the admissibility condition}: 
\begin{equation}
C_\psi = \|\psi \|^{-2} \int_{g\in G} |\bra\psi|U(g)\psi \ket|^2 d\mu(g) < \infty
,
\label{adc}
\end{equation}
where $d\mu(g)$ is the left-invariant Haar measure on $G$.

Then, any vector $|\phi\ket$ of a Hilbert space $\cH$ can be 
represented in the form:
\begin{equation}
|\phi\ket = C_\psi^{-1} \int_G |U(g)\psi\ket d\mu(g)\bra\psi U^*(g)|\phi\ket,
\label{pu}
\end{equation}
The equation \eqref{pu} is also known as {\em the partition of the unity} 
with respect to a Lie group $G$ and is often written in the form 
$$ \hat 1 = C_\psi^{-1} \int_G |U(g)\psi\ket d\mu(g)\bra\psi U^*(g)|.$$
The basic vector $\psi$ used to construct decomposition \eqref{pu} is 
often called {\em fiducial vector}, or {\em basic wavelet}. 

\subsection{Translation group: Fourier transform}
The most familiar case of the unity partition is the decomposition with 
respect to the momentum eigenstates 
\begin{equation}
|\phi\ket = \int |k\ket dk \bra k |\phi\ket.
\label{puk}
\end{equation} 
In the later case $G$ is the group of translations and 
the Haar measure is simply $dk$. There is no need in explicit notation 
of any fiducial vector $\psi$ there, because the group of translations 
is Abelian and the representation $U(k)$ is just a mapping between the 
vectors $k\in\R^n$ and the eigenvectors of the momentum operator. However, 
for the case of other Lie groups we have to put $\psi$ explicitly. 
Of course, the final physical results of the theory should be independent 
on fiducial vector $\psi$.
\subsection{Affine group: Wavelet transform}
For the case of affine group \eqref{ag}, $x,b\in\R^n$, with the $SO_n$ 
rotations dropped for simplicity, the left-invariant 
Haar measure is $d\mu(a,b) = a^{-n-1}da db$, the representation induced 
by a basic wavelet ($\psi(x)\in L^2(\R^n)$ for definiteness) is 
$U(g)=a^{-n/2}\psi((x-b)/a))$. So,
\begin{equation}
\begin{array}{lcl}
\phi_a(b) &=& \int  a^{-n/2}\bar\psi\left(\frac{x-b}{a}\right)\phi(x)d^nx, \\
\phi(x)&=& C_\psi^{-1}\int a^{-n/2}\psi\left(\frac{x-b}{a}\right)\phi_a(b)
\frac{dadb}{a^{n+1}},
\end{array} 
\label{wtl2}
\end{equation} 
where 
\begin{equation}
C_\psi = \int \frac{|\hat\psi(k)|^2}{|k|}d^nk.
\label{adcf}
\end{equation}
See e.g. \cite{Chui} for detailed explanation. 

\section{$\phi^4$ model on the affine group} 
Let us turn to the fourth power interaction model with the (Euclidean) 
action functional 
\begin{equation}
\begin{array}{lcl}
S[\phi] &=& \frac{1}{2}\int \phi(x_1)D(x_1,x_2)\phi(x_2)dx_1dx_2  \\
        &+& \frac{1}{4!}\int V(x_1,x_2,x_3,x_4)\phi(x_1)\phi(x_2)\phi(x_3)
          \phi(x_4)dx_1dx_2 dx_3dx_4
\label{Sphi}
\end{array}
\end{equation}
Using the notation 
$$U(g)|\psi\ket \equiv |g,\psi\ket, \quad \bra\phi|g,\psi\ket \equiv \phi(g),
\quad \bra g_1,\psi|D|g_2,\psi\ket \equiv D(g_1,g_2)
$$
we can rewrite the generating functional \eqref{gf} for the field theory 
with action \eqref{Sphi} in the form 
\begin{equation}
\begin{array}{lcl}
Z_G[J] &=& \int \cD\phi(g) \exp\Bigl(
-\frac{1}{2}\int_G \phi(g_1)D(g_1,g_2)\phi(g_2)d\mu(g_1) d\mu(g_2) \\
&-&\frac{\lambda_0}{4!}\int_G \tilde V(g_1,g_2,g_3,g_4)
\phi(g_1)\phi(g_2)\phi(g_3)\phi(g_4)
d\mu(g_1) d\mu(g_2)d\mu(g_3) d\mu(g_4)  \\
&+& \int_G J(g)\phi(g)d\mu(g)
\Bigr),
\end{array}
\label{gfi4}
\end{equation}
where $\tilde V(g_1,g_2,g_3,g_4)$ is the result of the application of 
the transform $$\tilde\phi(g) := \int\overline{U(g)\psi(x)} \phi(x)dx$$ 
to $V(x_1,x_2,x_3,x_4)$ in all arguments $x_1,x_2,x_3,x_4$.

Let us turn to the particular case of the affine group \eqref{ag}. 
The restriction imposed by the admissibility condition \eqref{adc} on 
the fiducial vector $\psi$ (the basic wavelet) is rather loose: 
Only the finiteness of the integral $C_\psi$ given by \eqref{adcf} is 
required.  
This practically implies only that $\int \psi(x)dx =0$ and that 
$\psi(x)$ has compact support. For this reason the wavelet transform 
\eqref{wtl2}
can be considered as a microscopic slice of the function $\phi(x)$ taken 
at a position $b$ and resolution $a$ with ``aperture'' $\psi$. Each 
particular aperture $\psi(x)$ of course has its own view, but the physical 
observable should be independent on it. In practical applications of 
WT very often either of the derivatives of the Gaussian 
$\psi_n(x) = (-1)^n d^n/dx^n e^{-x^2/2}$ is used, but for the purpose of 
the present paper only the admissibility condition is important but not 
the shape of $\psi(x)$. 

So, for the case of decomposition of scalar free field in $\R^n$ 
with respect to affine group, the inverse free field propagator matrix 
element is 
\begin{eqnarray*}
\bra a_1,b_1;\psi | D | a_2, b_2; \psi\ket = \int d^nx (a_1a_2)^{-\frac{n}{2}}
\bar\psi \left( \frac{x-b_1}{a_1} \right)  D \psi\left(\frac{x-b_2}{a_2}\right) \\
= \int \dk{k}{n} e^{ik(b_1-b_2)} (a_1a_2)^{\frac{n}{2}} \overline{\hat \psi(a_1 k)}
(k^2+m^2) \hat \psi(a_2 k) \\
\equiv \int \dk{k}{n} e^{ik(b_1-b_2)} D(a_1,a_2,k).
\end{eqnarray*} 
Assuming the homogeneity of the free field in space coordinate, i.e. that 
matrix elements depend only on the differences $(b_1-b_2)$ of the positions, 
but not the positions themselves, we can use  $(a,k)$ representation:
\begin{eqnarray}
\nonumber D(a_1,a_2,k) &=& a_1^{n/2} \overline {\hat \psi(a_1 k)} (k^2+m^2)
                 a_2^{n/2}\hat \psi(a_2 k) \\
D^{-1}(a_1,a_2,k) &=& a_1^{n/2} \overline {\hat \psi(a_1 k)} 
                  \left( \frac{1}{k^2+m^2} \right)
                 a_2^{n/2}\hat \psi(a_2 k)\\
\nonumber d\mu(a,k)       &=& \dk{k}{n}\da{a}{n}.
\end{eqnarray}
So, we have the same diagram technique as usual, but with extra 
``wavelet'' term $a^{n/2}\hat \psi(a k)$ term on each line and the integration 
over $d\mu(a,k)$ instead of $dk$. 

Concerning the Lorenz covariance of the resulting theory (i.e. invariance 
under rotations, since Euclidean version of the theory is considered), the  
introduction of the new scale variable $a$, practically means that instead 
one scalar field $\phi(x)$, we have to deal with a collection of fields 
labeled by the resolution parameter $\{ \phi_a(x) \}_a$. For each of them 
the invariance under rotations and translations holds of course. The things 
are so simple only if we assume quantization = functional integration 
in the space of numeric-valued functions only, without paying any special 
attention to possible commutation relations $[\phi_{a_1}(x),\phi_{a_2}(y)]$.
What happens for the case of operator-valued functions is not clear enough 
\cite{Federbush}.   

Now, turning back to the coordinate representation \eqref{wtl2}, where 
$a$ is the resolution (``window width'') and recalling the power law 
dependence \eqref{ws} obtained by Wilson expansion, we can define the 
$\phi^4$ model on affine group, with the coupling constant 
dependent on scale. The simplest case of fourth power interaction 
of this type is
\begin{equation} 
V_{int} =\int \frac{\lambda(a)}{4!} \phi^4_a(b) d\mu(a,b),
\quad  \lambda(a)\sim a^\nu.
\label{vint}
\end{equation}

The one-loop order contribution to the Green function $G_2$ 
in the theory with interacion \eqref{vint}
can be easily evaluated 
(for isotropic wavelet $\hat\psi(\mathbf{k})=\hat\psi(k)$, otherwise 
the constants will be different) by integration over a scalar variable 
$z = ak$: 
\begin{equation}
\int \frac{a^\nu a^n |\hat\psi(ak)|^2 }{k^2+m^2}\dk{k}{n}\da{a}{n}  
= C_\psi^{(\nu)}  \int \dk{k}{n} \frac{k^{-\nu}}{k^2+m^2}, 
\end{equation} 
where $$C_\psi^{(\nu)}=\int |\hat\psi(\bvec{k})|^2 k^{\nu-1}dk.$$ 
Therefore, there are no ultra-violet divergences for $\nu>n-2$. 

In the next orders of perturbation expansion each vertex will contribute 
$k^{-\nu}$ to the formal divergence degree of each diagram. This is quite 
natural from dimensional consideration, for $a$ - is a ``scale'' (window 
width), and $k$ is ``inverse scale''.  
So far, for the $\phi^N$ theory 
in $n$ dimensions a diagram with $E$ external lines and $V$ vertexes 
has a formal divergence degree 
\begin{equation}
D = n + E\left({1-\frac{n}{2}}\right) 
+ V\left({\frac{n}{2}(N-2)-N-\nu}\right). 
\label{dd}
\end{equation} 
\section{Conclusion} 
Of course, the power law behavior of the coupling constant 
$\lambda(a)=a^\nu$ is not quite realistic. It seems more 
natural if $\lambda$ vanish outside a limited domain of scales. 
The physical meaning of considering a 
field theory on the affine group seems more important. Doing 
so we acquire two parameters: the {\em coordinate} $b$ and the 
{\em resolution} 
$a$. The former is present in any field theory, but the later is 
not. Such model can be considered as a continuous counterpart of 
lattice theory, but now we consider the grid size, or scale, as a physical 
parameter of interaction and there is no need to get rid of it 
at the end of calculations.  
\vskip\baselineskip
\centerline{***}
The author is grateful to the Department of Mathematics, University 
of Alberta, where this research was partially done, for financial 
support and stimulating interest and also to Dr. T.Gannon 
and Prof. V.B.Priezzhev for useful discussions. 

\end{document}